\documentclass[12pt,twoside]{article}
\usepackage{fleqn,espcrc1,epsfig}

\newcommand{\AmS}{{\protect\the\textfont2
  A\kern-.1667em\lower.5ex\hbox{M}\kern-.125emS}}
\newcommand{\J}{J/\Psi}

\title{\vspace*{-1.7cm}
Critical Review Of Quark Gluon Plasma Signals 
\thanks{This work was supported by Deutsche 
Forschungsgemeinschaft (DFG), 
Bundesministerium f\"ur Bildung und Forschung (BMBF), 
Gesellschaft f\"ur Schwerionenforschung
(GSI) and 
Graduiertenkolleg Theoretische und Experimentelle Schwerionenphysik.
One of us (LG) is supported by the Josef Buchmann Foundation. }}
\author{D. Zschiesche\address{Institut f\"ur Theoretische Physik,
 D-60054 Frankfurt am Main, Germany},
        L. Gerland${}^{\rm{a}}$,
        S. Schramm${}^{\rm{a}}$,
J. Schaffner-Bielich
\address{Riken BNL Research Center, Brookhaven National Lab, Upton,
New York 11973}, H. St\"ocker${}^{\rm{a}}$  
\footnote{Invited Speaker}
, and W. Greiner${}^{\rm{a}}$}
\date{\today}
\begin{document}
\maketitle
Compelling evidence for a
new form of matter has been claimed to be formed in 
Pb+Pb collisions at SPS. We critically 
review two suggested signatures for this new state of matter:
First the suppression of the J/$\Psi$,  
which should be strongly suppressed 
in the QGP by two different mechanisms, the
color-screening~\cite{MaT86} and  
the QCD-photoeffect~\cite{KhD96}.
Secondly  the  
measured particle, in particular strange hadronic, ratios 
might signal the freeze-out from a 
quark-gluon phase. 
\section{J/$\Psi$ suppression}
The QCD factorization theorem is used to evaluate the PQCD cross sections of
heavy quarkonium interactions with ordinary hadrons. However, the charmonium
states (here denoted $X$) are not sufficiently small to ignore nonperturbative
QCD physics.
Thus, we evaluate the nonperturbative QCD contribution to the cross sections
of charmonium-nucleon interaction by using an interpolation between
known cross sections~\cite{ger}. 
The $\J$-$N$ cross section evaluated in this paper is
in reasonable agreement with SLAC data \cite{slac}.

Indeed, the $A$-dependence of the $\J$ production studied at SLAC
at $E_{inc} \sim 20$ GeV exhibits a significant absorption effect \cite{slac}
leading to $\sigma_{abs}(\J$-$N)= 3.5 \pm 0.8$ mb.
It was demonstrated~\cite{farrar} that, in the kinematic region at SLAC,
the color coherence effects are still small on the internucleon
scale for the formation of $\J$'s.
So, in contrast to the findings at higher energies, at intermediate
energies this process measures the {\it genuine} $\J$-$N$
interaction cross section at energies of $\sim $ 15-20 GeV \cite{farrar}.

To evaluate the nonperturbative QCD contribution we use an
interpolation formula~\cite{ger} for the dependence of the cross section
on the transverse size $b$ of a quark-gluon configuration
Three reference points are used to fix our parametrization of
the cross sections (cf. Tab.~\ref{meanb}).
The $X$-$N$ cross sections is calculated via:
$
\sigma=\int \sigma(b)\cdot |\Psi (x,y,z)|^2 {\rm d}x\, {\rm d}y\,
{\rm d}z 
$,
where $\Psi (x,y,z)$ is the charmonium wave function. In our calculations
we use the wave functions from a non-relativistic charmonium model 
(see~\cite{werner}).%

\begin{table}[h]
\vspace{-.5cm}
\begin{center}
\begin{tabular}{|c|c|c|c|c|}
\hline
$c\overline{c}$-state & J/$\Psi$ & $\Psi'$ & $\chi_{c10}$ & $\chi_{c11}$\\
\hline
$\sigma$ (mb) & 3.62 & 20.0 & 6.82 & 15.9 \\
\hline
\end{tabular}
\end{center} 
\vspace{-.5cm}
\caption
{\label{meanb}\small{The total quarkonium-nucleon cross sections $\sigma$.
For the $\chi$ two values arise, due to the spin dependent wave functions
($lm=10,11$).}}
\end{table}

We follow the analysis of~\cite{kharzeev} to evaluate
the fraction of $\J$'s (in $pp$ collisions) that come from the decays of
the $\chi$ and $\Psi'$. So, the
suppression factor $S$ of $\J$'s produced in the nuclear medium is
calculated as:\\
$
S=0.6\cdot ( 0.92\cdot S^{\J}+0.08\cdot S^{\Psi'})+0.4\cdot S^{\chi}
$.
Here $S^X$ are the respective suppression factors of the different
pure charmonium states $X$ in nuclear
matter. The $S^X$ are for minimum bias $pA$ collisions within the
semiclassical approximation (cf.~\cite{hufner}).

The charmonium states are produced as small
configurations, then they evolve to
their full size.
Therefore, if the formation length of the charmonium states, $l_f$, becomes
larger than the average internucleon distance,
 one has to take into account the evolution of
the cross sections with the distance from the production point~\cite{farrar}.

The formation length of the $\J$ is given by 
$l_f\approx \frac{2p}{m^2_{\Psi'}- m^2_{\J}}$, where $p$ is the momentum of
the $\J$ in the rest frame of the target. For a $\J$ produced at midrapidity 
at SPS energies, this yields $l_f\approx 3$ fm.
Due to the lack of better knowledge, we use the same $l_f\approx 3$ fm for the
$\chi$. For the $\Psi'$ we use $l_f\approx 6$ fm, 
because it is not a small object, but has the size of
a normal hadron, i.e. the pion. For $E_{lab}=800$ AGeV we get a factor
of two for the formation lengths due to the larger Lorentz factor.

\begin{figure}[h]
\centerline{\parbox[b]{6cm}{\epsfxsize=8.5cm
\vspace*{-2.5cm}
\epsfbox{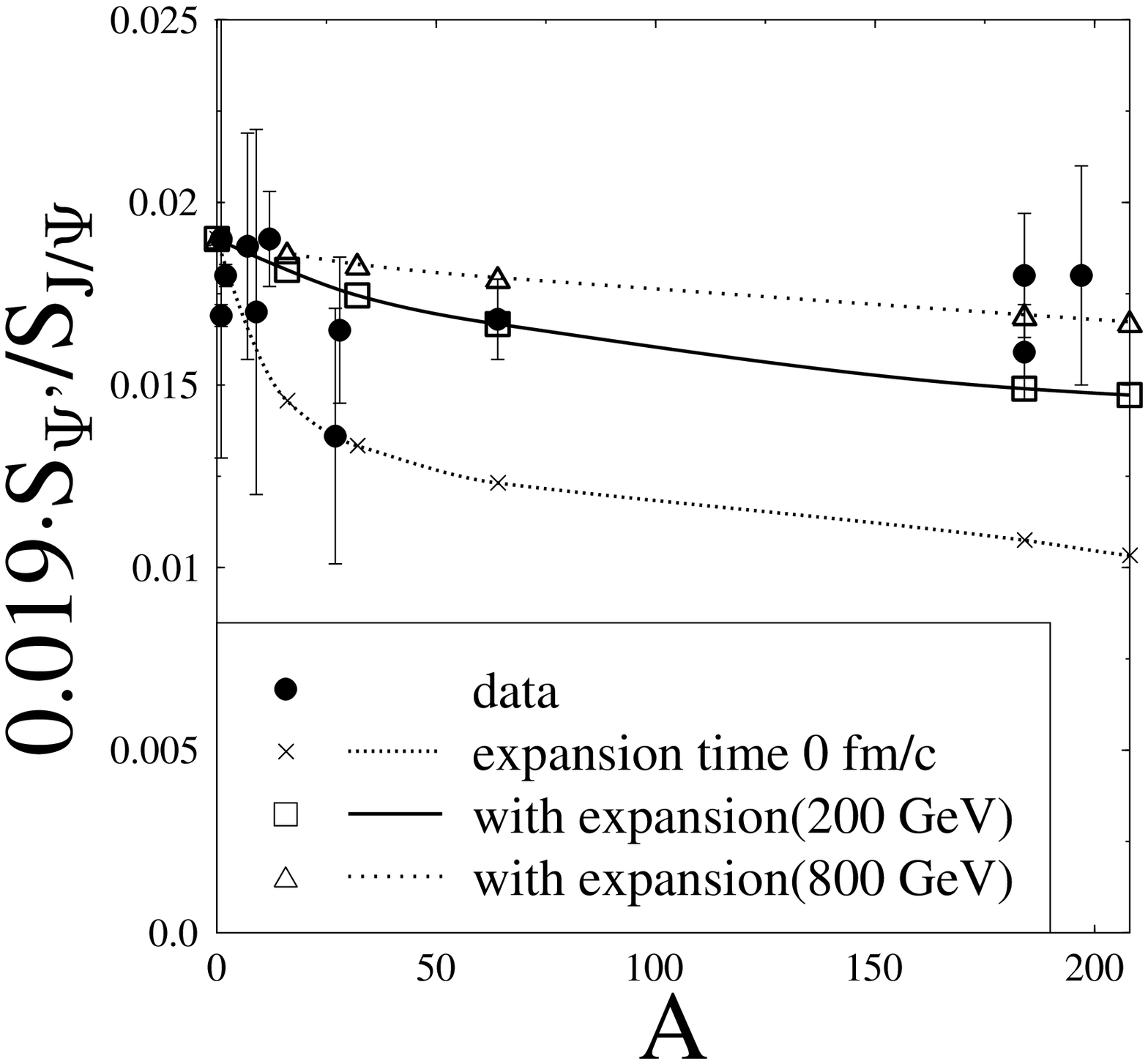}}\hfill
\vspace{-1cm}
\parbox[b]{6cm}{\caption{\small}
The ratio $0.019\cdot S_{\Psi'} / S_{\J}$ is shown in $pA$ (crosses)
in comparison to the data (circles). The squares and the triangles shows the
ratio calculated with the expansion of small wave packages
(see text).
\vspace{3cm}\label{ratio}}}
\vspace*{-1cm}
\end{figure}

However, this has a large impact on the
$\Psi'$ to $\J$-ratio depicted in Fig.~\ref{ratio}, which
shows the ratio $0.019\cdot S_{\Psi'} / S_{\J}$ calculated with (squares
(200 GeV) and triangles (800 GeV))
and without (crosses) expansion. The factor 0.019 is the measured value in
$pp$ collisions, because the experiments do not measure the calculated value
$S_{\Psi'} / S_{\J}$ but
$\frac{B_{\mu\mu}\sigma(\Psi')}{B_{\mu\mu}\sigma(\J)}$.
$B_{\mu\mu}$ are the branching ratios for $\J ,\,\Psi'\rightarrow\mu\mu$.

The calculations which take into
account the expansion of small wave packages show better agreement
with the data (circles) (taken from~\cite{carlos}) than the calculation
without expansion time, i.e. with immediate $\J$ formation, $l_f=0$.   
We calculated this effect both at $E_{lab}=200$ AGeV and 800 AGeV.     
The data have been measured at different energies ($E_{lab}$ = 200, 300, 400,
450, 800 GeV and $\sqrt{s}$ = 63 GeV). One can see that this ratio is nearly
constant in the kinematical region of the data, but it decreases at smaller
momentum (e.g. $E_{lab}=200$ AGeV and $y<0$) due to the larger cross section 
of the $\Psi'$.

However, the P-states yield two vastly different
cross sections (see Tab.~\ref{meanb}) for $\chi_{10}$ and $\chi_{11}$,
respectively. This leads to a higher absorption rate of the $\chi_{11}$ as
compared to the $\chi_{10}$. This new form of color filtering is predicted
also for the corresponding states of other hadrons; e.g. for the bottomium
states which are proposed as contrast signals to the $\J$'s at RHIC and LHC!

Furthermore it is important to also take into account
comoving mesons. 
Therefore we use the UrQMD model \cite{bass98,spieles99}. 
Particles
produced by string fragmentation are not allowed to interact with other
 hadrons --
in particular with a charmonium state -- within their formation time 
(on average, $\tau_F\approx 1$ fm/c). However,
leading hadrons are allowed to interact with a reduced cross section even
within their formation time . 
The reduction factor is 1/2 for mesons which
contain a leading constituent quark from an incident nucleon and 2/3 for 
baryons which contain a leading diquark.

\begin{figure}[h]
\centerline{\parbox[b]{6cm}{\epsfxsize=8.5cm
\vspace*{-1.5cm}
\epsfbox{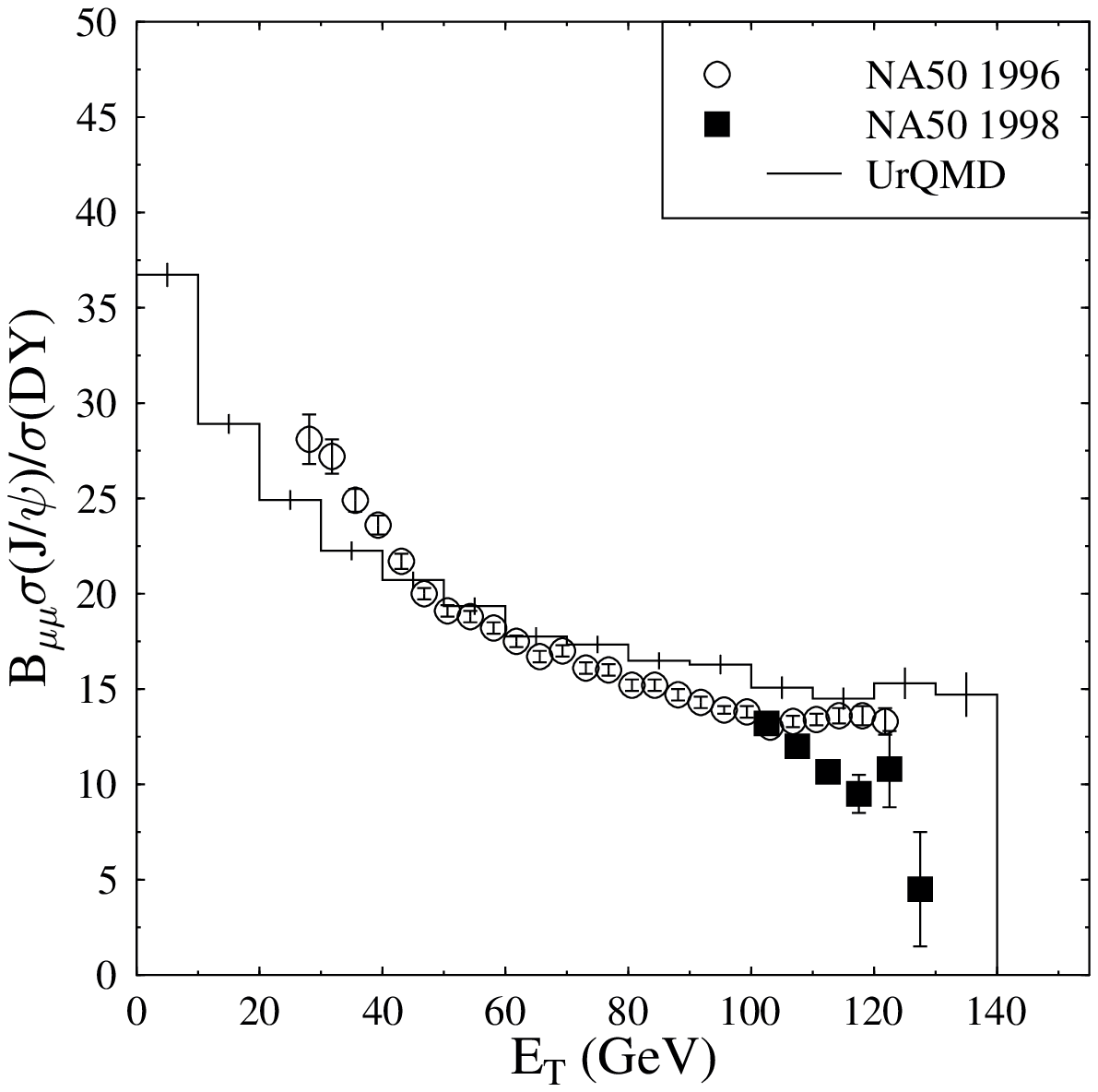}}\hfill
\parbox[b]{6cm}{\caption{\small}
The ratio of $J/\psi$ to Drell-Yan production as a function of
$E_T$ for Pb+Pb at 160~GeV (see text).
\vspace{4cm}\label{psidyet}}}
\vspace*{-1cm}
\end{figure}

Figure~\ref{psidyet} shows the $J/\psi$ to Drell-Yan ratio
as a function of $E_T$ for Pb+Pb interactions at 160~GeV compared to
the NA50 data \cite{romana,na50neu}.
The normalization of $B_{\mu\mu}\sigma(J/\psi)/\sigma({\rm DY})=46$ in $pp$
interactions at 200~GeV has been
fit  to S+U data within a geometrical
model \protect\cite{kharzeev}.
The application of this value to our analysis is not arbitrary:
the model of Ref.~\protect\cite{kharzeev} renders
the identical $E_T$-integrated $J/\psi$ survival probability, $S=0.49$,
as the UrQMD calculation for this system.
An additional factor of 1.25
\protect\cite{reviewvogt} has been applied to the Pb+Pb calculation
in order to account for the lower energy, 160 GeV, since the
$J/\psi$ and Drell-Yan cross sections have different energy and isospin
dependencies.

The gross features of the $E_T$ dependence of the $J/\psi$ to Drell-Yan
ratio are reasonably well described by the model calculation.
No discontinuities in the shape of the ratio as a function of $E_T$
are predicted by the
simulation. The new high $E_T$ data \cite{na50neu} decreases stronger
than the calculation. This could be caused by underestimated 
fluctuations of the 
multiplicity of secondaries in the UrQMD model. This occurs, since
high $E_T$-values are a trigger for very central events with a
secondary multiplicity larger than in average \cite{cap00}. 

\section{Particle production}
Ideal gas model calculations have been used for a long time to
calculate particle production in relativistic heavy ion collisions, e.g.
\cite {hahn86,hahn88,brau99,raf99,bec00,yen98}. 
Fitting the particle ratios as obtained 
from those ideal gas calculations to the experimental measured ratios
at  SIS, AGS and SPS for different energies and different colliding
systems yields a curve of  chemical freeze-out in the $T-\mu$ plane.
Now the question 
arises, how much the deduced temperature and chemical potentials 
depend on the model employed. Especially the influence of changing hadron
masses and effective potentials should be investigated, as has been
done for example in \cite{stoe78,thei83,scha91,springer}.
This is of special importance for the quest of a signal of the formation of a
 deconfined phase, i.e. the quark-gluon plasma. As deduced from lattice data 
\cite{kar98}, the critical temperature for the onset of a deconfined phase
 coincides with that of a chirally restored phase. Chiral effective models of
 QCD therefore can be utilized to give important insights on signals from a
 quark-gluon plasma formed in heavy-ion collisions.
  
\begin{figure}[hb]
\centerline{\parbox[b]{6cm}{\epsfxsize=8.5cm
\vspace*{-1.5cm}
\epsfbox{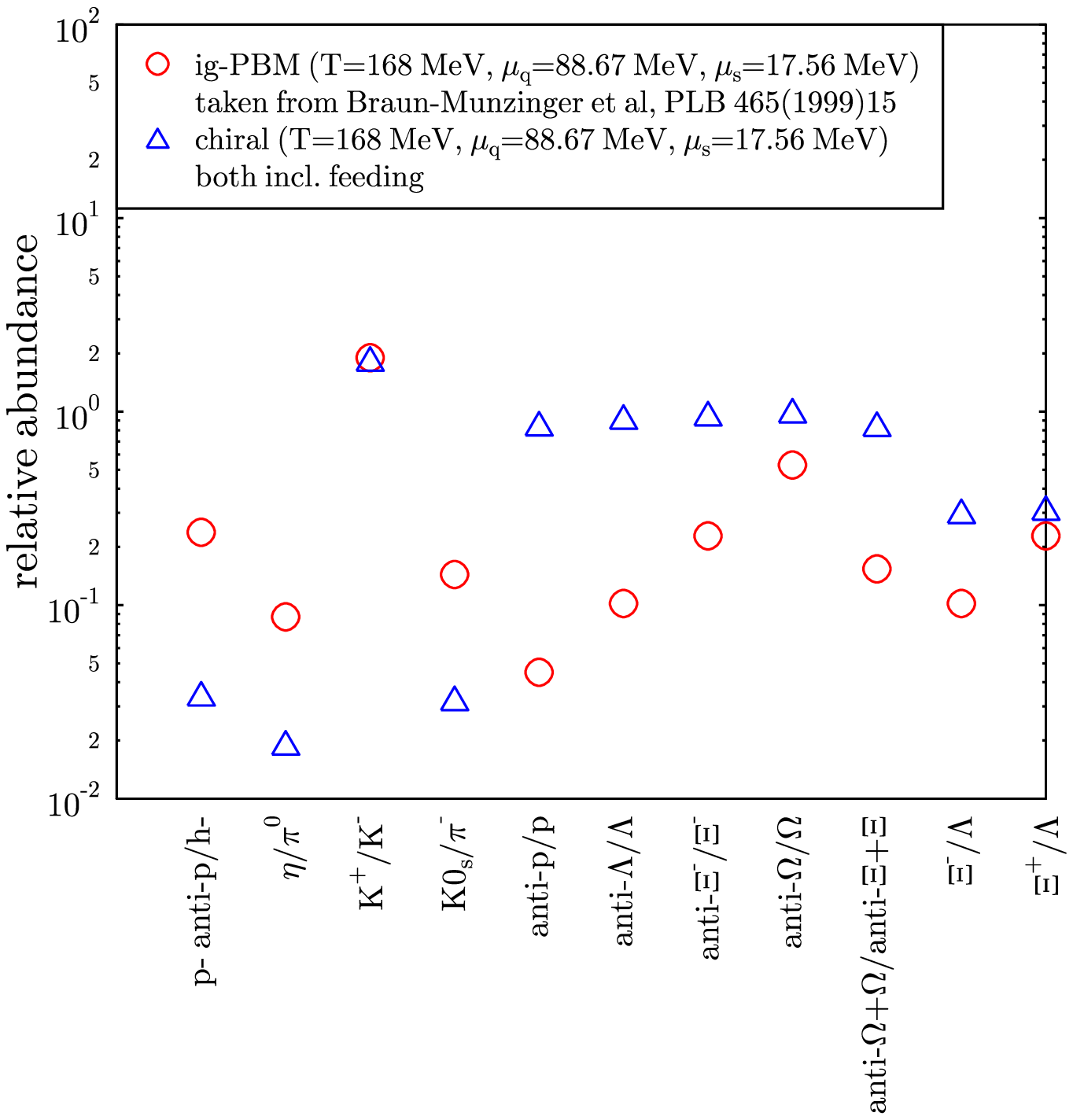}}\hfill
\parbox[b]{6cm}{\caption{\small}Comparison of a chiral calculation with 
\protect\cite{brau99} for $T=168 \rm{MeV}, \mu_q = 88.67 \rm{MeV}$. The huge
differences result from the dropping baryonic masses, depending
on the strangeness content of the particles, and from the change of the
effective potentials in the chirally restored phase.
\vspace{3cm}\label{igchiral}}}
\vspace*{-1cm}
\end{figure}

\begin{figure}[h]
\centerline{\parbox[b]{6cm}{\epsfxsize=10cm
\vspace*{-1.5cm}
\epsfbox{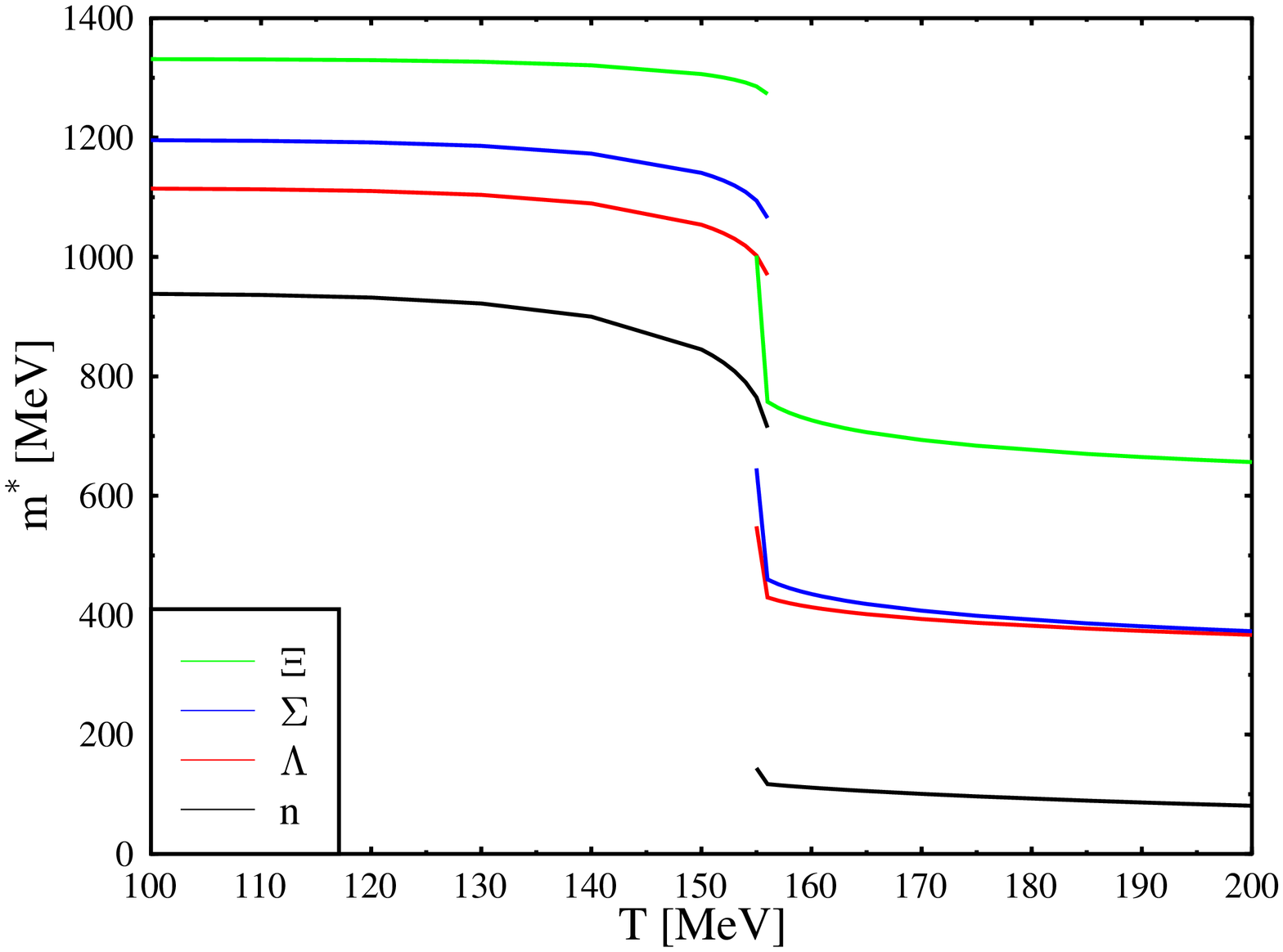}}\hfill
\parbox[b]{6cm}{\caption{\small}
Baryon masses as function of temperature for vanishing
chemical potential. One sees that a phase-transition occurs at 
$T_c \approx 150 MeV$. There the masses drop significantly and then
nearly saturate. The baryon mass above $T_c$ scales with 
the strangeness content of the corresponding baryon.
\vspace{1cm}\label{bmasses}}}
\vspace*{-1cm}
\end{figure}

Therefore we compare experimental measurements for Pb+Pb collisions at
SPS with the ideal gas calculations and 
results obtained from a chiral SU(3) model \cite{paper3,springer}. 
This effective hadronic model 
predicts a chiral phase transition at $T \approx 150 MeV $.
Furthermore the
model predicts changing hadronic masses and effective chemical
potentials, due to strong scalar and vector fields in hot and dense
hadronic matter, which are constrained by chiral symmetry from the 
QCD Lagrangean.\\
In \cite{brau99} 
the ideal gas model was fitted to particle ratios measured in Pb+Pb
collisions at SPS. The lowest 
$\chi^2$ is obtained for $T=168 \rm{MeV}$ and $\mu_q= 88.67 \rm{MeV}$. Using
these values as input for the chiral model leads to dramatic changes.
As can be seen in Figure \ref{igchiral}. 
There are two main reasons
for the strong deviations. First, since the chosen temperature lies
above the chiral phase transition temperature of the model, 
the effective masses of
the baryons are lowered dramatically
(see Figure \ref{bmasses}).
The second
reason are the strongly changed effective potentials. While  the chemical 
potential of the proton with
the chosen parameters is $\mu_p = 266 \rm{MeV}$ in the
ideal gas model, the effective chemical
potenial of the proton in the chiral model is only $\mu_p \approx 8 \rm{MeV}$,
due to the reduction by
the vector-field. Such a small effective chemical
potential leads to nearly equal particle-antiparticle numbers and
therefore to tremendeously enhanced particle-antiparticle ratios.
Obviously, the freeze-out temperature and chemical potential have to be 
readjusted to account for the in-medium effects of the hadrons in the chiral 
model.

\begin{figure}[h]
\centerline{\parbox[b]{6cm}{\epsfxsize=10cm
\vspace*{-1.5cm}
\epsfbox{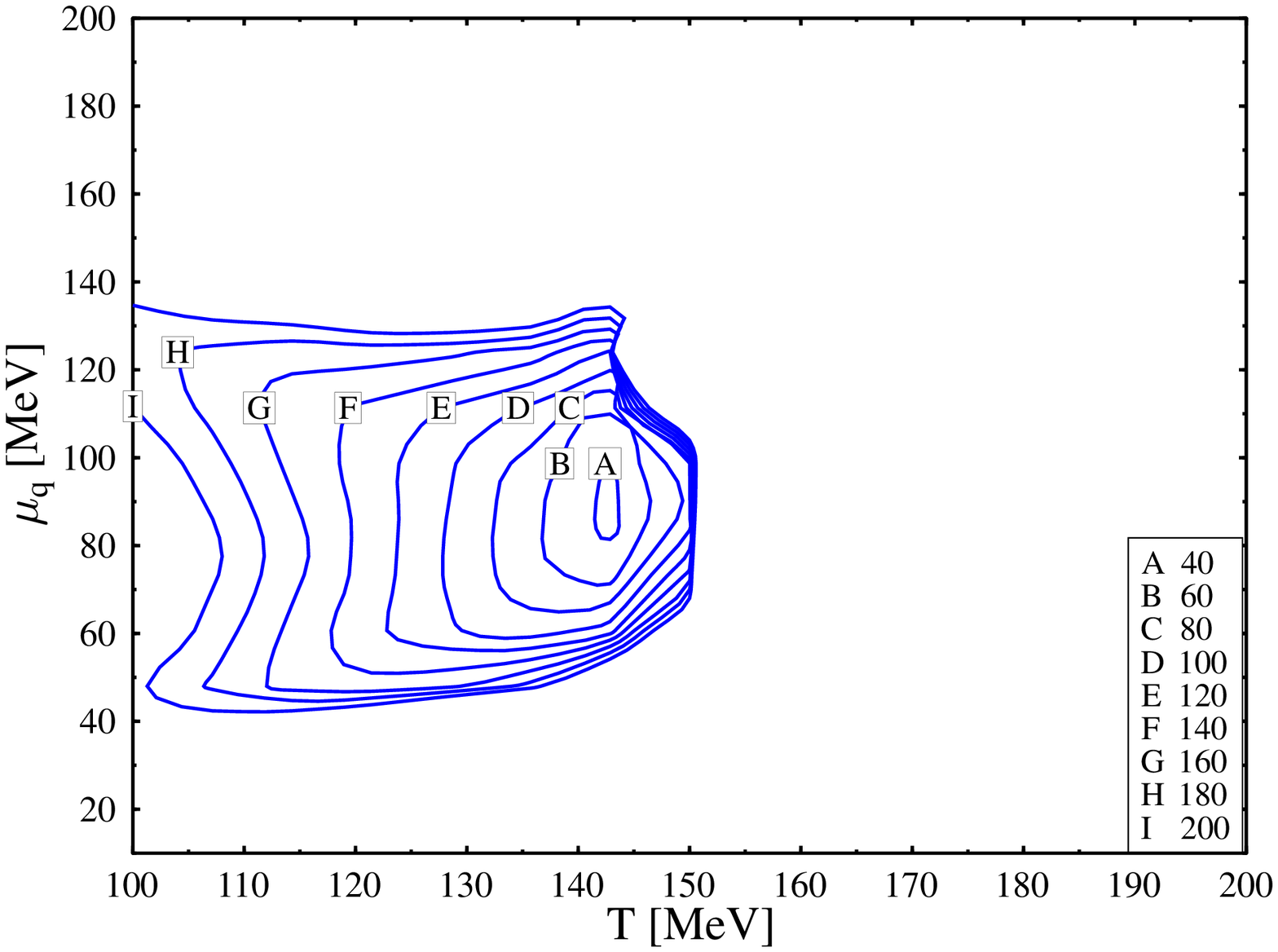}}\hfill
\parbox[b]{6cm}{\caption{\small}
$\chi^2$ for chiral model, 
data taken from \protect\cite{brau99}. The best fit
parameters are $T = 144 MeV$ and $\mu_q \approx 95 MeV$. $\mu_q$ is
chosen such that $f_s=0$. For $T>T_c$ no agreement with data 
from CERN SPS lead on lead collisions can be obtained.
\vspace{2cm}\label{chi2chiral}}}
\vspace*{-1cm}
\end{figure}

\begin{figure}[h]
\centerline{\parbox[b]{6cm}{\epsfxsize=8cm
\vspace*{-1.5cm}
\epsfbox{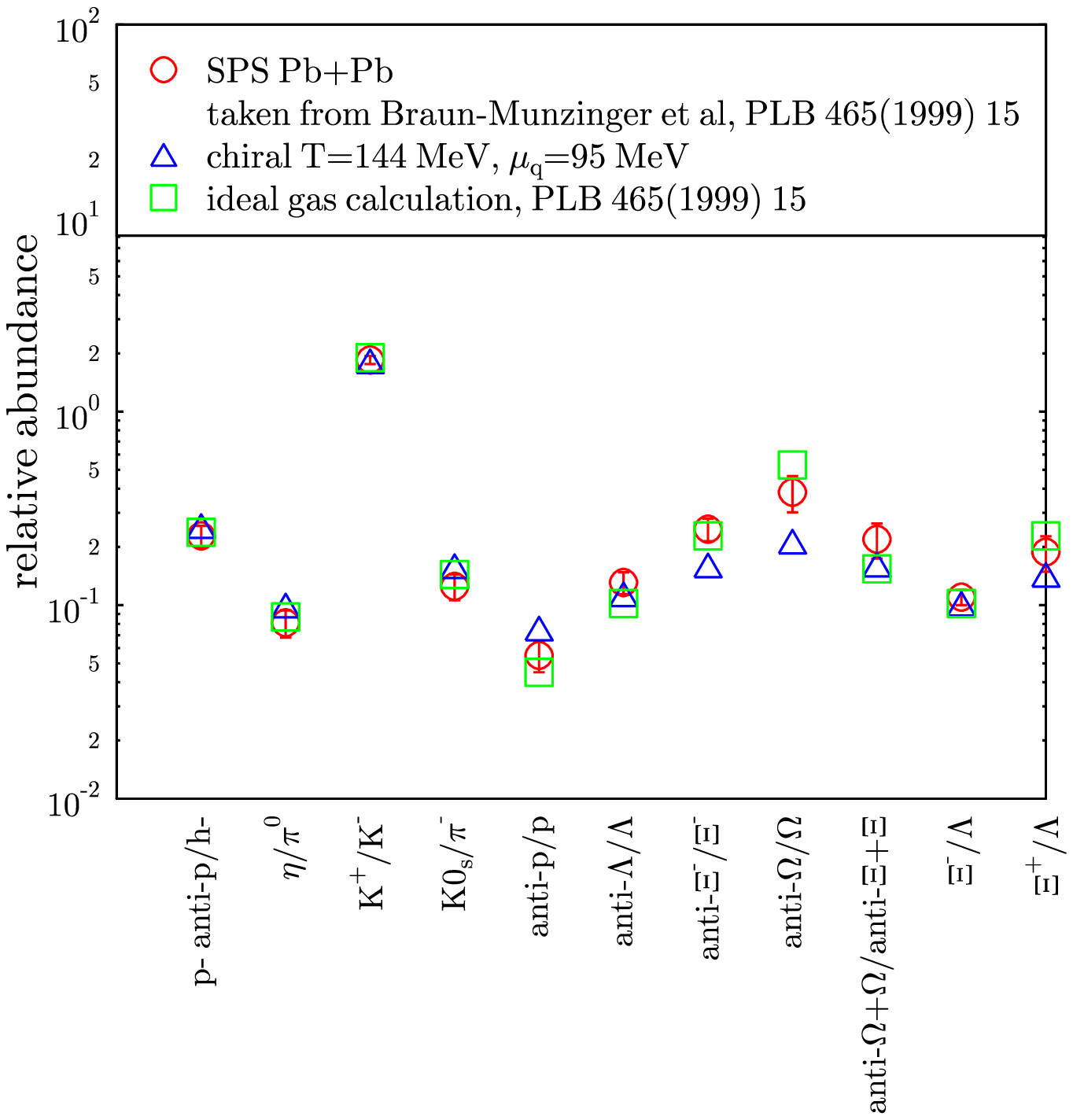}}\hfill
\parbox[b]{6cm}{\caption{\small}
Comparison of chiral model for $T=144$ MeV 
compared to data and ideal gas results from 
\protect\cite{brau99}.
\vspace{3cm}\label{chiralex}}}
\vspace*{-1cm}
\end{figure}

We call the best fit the parameter set that gives a minimum in 
the value of $\chi^2$, with
\begin{equation}
\chi^2 = \sum_i \frac {\left(r_i^{exp} - r_i^{model}\right)^2}{\sigma_i^2}.
\end{equation}
Here $r_i^{exp}$ is the experimental ratio, $r_i^{model}$ is the ratio
calculated in the model and $\sigma_i$ represents the error in the
experimental data points as quoted in \cite{brau99}.
We included the following ratios in the fit procedure:
$\frac{\bar{p}-p}{h^-}$, $\frac{\eta}{\pi^0}$,
$\frac{K^+}{K^-}$,$\frac {K^0_s}{\pi^-}$,
$\frac{\bar{p}}{p},\frac{\bar{\Lambda}}{\Lambda}$,
$\frac{\bar{\Xi}}{\Xi}$,$\frac{\bar{\Omega}}{\Omega}$,
$\frac{\bar{\Omega}+\Omega}{\bar{\Xi^-}+\Xi^-}$,
$\frac{\bar{\Xi^-}}{\Lambda}$,$\frac{\Xi^+}{\Lambda}$.
The resulting values of $\chi^2$ for different $T-\mu$ pairs are shown 
in figure \ref{chi2chiral}.
 In all calculations $\mu_s$ was chosen
such that the overall net strangeness $f_s$ is zero. 
The best values for 
the parameters are $T = 144 \rm{MeV}$ and $\mu_q \approx 95
\rm{MeV}$. 
While the
value of the chemical potential does not change much compared to the
ideal gas calculation, the value of the temperature is lowered by
more than 20 MeV. Furthermore Figure \ref{chi2chiral} shows,
that the dropping effective masses
and the reduction of the effective chemical potential make the
reproduction of experimentally measured particle ratios as seen at CERN's SPS 
within this
model impossible for $T > T_c$.
Only for temperatures below $T_c$ we obtain
reasonable fits.
Using the best fit parameters we compare the particle ratios, as
calculated in the chiral model, with the ideal gas calculation of 
\cite{brau99} and data as compiled in \cite{brau99}. This is shown in figure
\ref{chiralex}. One obtains a satisfactory agreement 
over several magnitudes. The $\chi^2$ values of the chiral model is 
$\chi^2_{chiral} = 26.5$. 
This is larger than the value in the ideal gas
model of \cite{brau99} ($\chi^2_{ig} = 13$). 
Note that in \cite{brau99}
weak decays are accounted for. If we use our ideal gas calculation without
feeding from weak decays we obtain only slightly changed best values
($T = 168, \mu_q = 82 MeV$) compared to \cite{brau99} and 
$\chi^2_{ig-FFM} = 21.6$. 
This shows that the chiral and ideal gas analysis using the same feeding
procedures yield comparable agreement with data concerning the value of 
$\chi^2$. Figure \ref{chiralex} shows that there is satisfactorily 
agreement between data and experiment.
We want to emphasize, that in
spite of the strong assumption of thermal and chemical equilibrium
the obtained values for $T$ and $\mu$
differ significantly depending on the underlying model, i.e. whether and
how effective masses and effective chemical potentials are accounted for.
In our model, we conclude that the observed particle ratios as measured at 
CERN´s SPS do not signal the freeze-out from a chirally restored quark-gluon 
phase. Note that we assume implicitly, that the particle ratios are 
determined by the medium effects and freeze out during the late stage 
expansion - no flavor changing collisions occur anymore, but the
hadrons can take the necessary energy to get onto their mass shall by
drawing energy from the fields.
Rescattering effects will alter our conclusion but are presumably small when 
the chemical potentials are frozen.


%



\bibliographystyle{prsty}

\end{document}